# Due "paradossi meccanici" della Collezione Storica degli Strumenti di Fisica dell'Università di Palermo

**Aurelio Agliolo Gallitto, Maria Casula, Daniela Cirrincione, Emilio Fiordilino, Filippo Mirabello, Francesca Taormina**

Dipartimento di Fisica e Chimica, Università degli Studi di Palermo, via Archirafi 36, 90123 Palermo, Italy

E-mail: aurelio.agliologallitto@unipa.it

**Riassunto.** Molti degli strumenti della Collezione Storica degli Strumenti di Fisica dell'Università di Palermo risalgono all'inizio dell'Ottocento, periodo in cui si cominciò a introdurre la Fisica sperimentale negli studi universitari usando strumenti dimostrativi e apparati sperimentali durante le lezioni in aula per illustrare le leggi della Fisica. Tra i vari strumenti della Collezione vi sono anche i cosiddetti "paradossi", strumenti con sorprendenti proprietà che sembrano non seguire le leggi della Fisica. In questo articolo analizzeremo due "paradossi meccanici" della Collezione e discuteremo del loro possibile uso didattico.

**Abstract.** Many instruments of the Historical Collection of the Physics Instruments of the University of Palermo date back to the early nineteenth century, when experimental Physics begun to be taught in university studies by using instruments and apparatuses in the classroom to illustrate the laws of Physics. Among the various instruments belonging to the Collection, there are also the so-called "paradoxes", instruments with surprising properties that do not seem to follow the laws of Physics. In this article we analyze two "mechanical paradoxes" of the Collection and discuss their possible educational use.

## 1. Introduzione

La Collezione Storica degli Strumenti di Fisica dell'Università di Palermo rappresenta un'importante testimonianza dello sviluppo della Fisica negli ultimi duecento anni [1,2], sia dal punto di vista didattico sia dal punto di vista della ricerca scientifica. Infatti, molti degli strumenti della Collezione risalgono all'inizio dell'Ottocento, periodo in cui si cominciò a introdurre la Fisica Sperimentale negli studi universitari usando strumenti dimostrativi e apparati sperimentali durante le lezioni in aula per illustrare le leggi della Fisica [3]. Tra i vari strumenti della Collezione vi sono anche i cosiddetti "paradossi", strumenti con sorprendenti proprietà che sembrano non seguire le leggi della Fisica.

A Palermo lo studio della Fisica sperimentale ebbe inizio alla fine del 1700 con padre Eliseo della Concezione, anche se la vera svolta si ebbe con l'abate Domenico Scinà, che ottenne nel 1811 la cattedra di Fisica Sperimentale alla Reale Università degli Studi di Palermo [4,5]. E proprio nell'inventario Scinà (antecedente al 1832) sono elencati due paradossi meccanici: il doppio cono e il cilindro impiombato [5]. La prima pagina dell'inventario Scinà è mostrata in figura 1.

Dal punto di vista didattico, per richiamare l'attenzione degli studenti e nello stesso tempo illustrare le leggi della fisica, si può far ricorso ai giochi scientifici [6-8] e ai paradossi [9-11]. Questi strumenti consentono di mostrare effetti sorprendenti che, nel caso dei paradossi, sembrano contraddire il senso comune. Tuttavia, l'attenta osservazione degli esperimenti rivela che i fenomeni sono perfettamente coerenti con le leggi della Fisica, superando l'iniziale senso di stupore.

In questo articolo, discuteremo come la Collezione Storica degli Strumenti di Fisica possa essere una risorsa da cui prendere idee per sviluppare attività laboratoriali ed exhibit rivolti sia a studenti dei primi anni del liceo sia al pubblico generico. In particolare, analizzeremo due paradossi meccanici della Collezione: il doppio cono e il cilindro impiombato. Prima, determineremo le condizioni geometriche a cui deve soddisfare il doppio cono per risalire sulla guida inclinata. Dopo, analizzeremo le condizioni di equilibrio statico del cilindro impiombato sopra il piano inclinato e determineremo la massima inclinazione che può avere il piano affinché il cilindro rimanga in equilibrio. Discuteremo quindi come questi due paradossi meccanici possono essere usati per stimolare l'interesse degli studenti nello studio della Fisica.

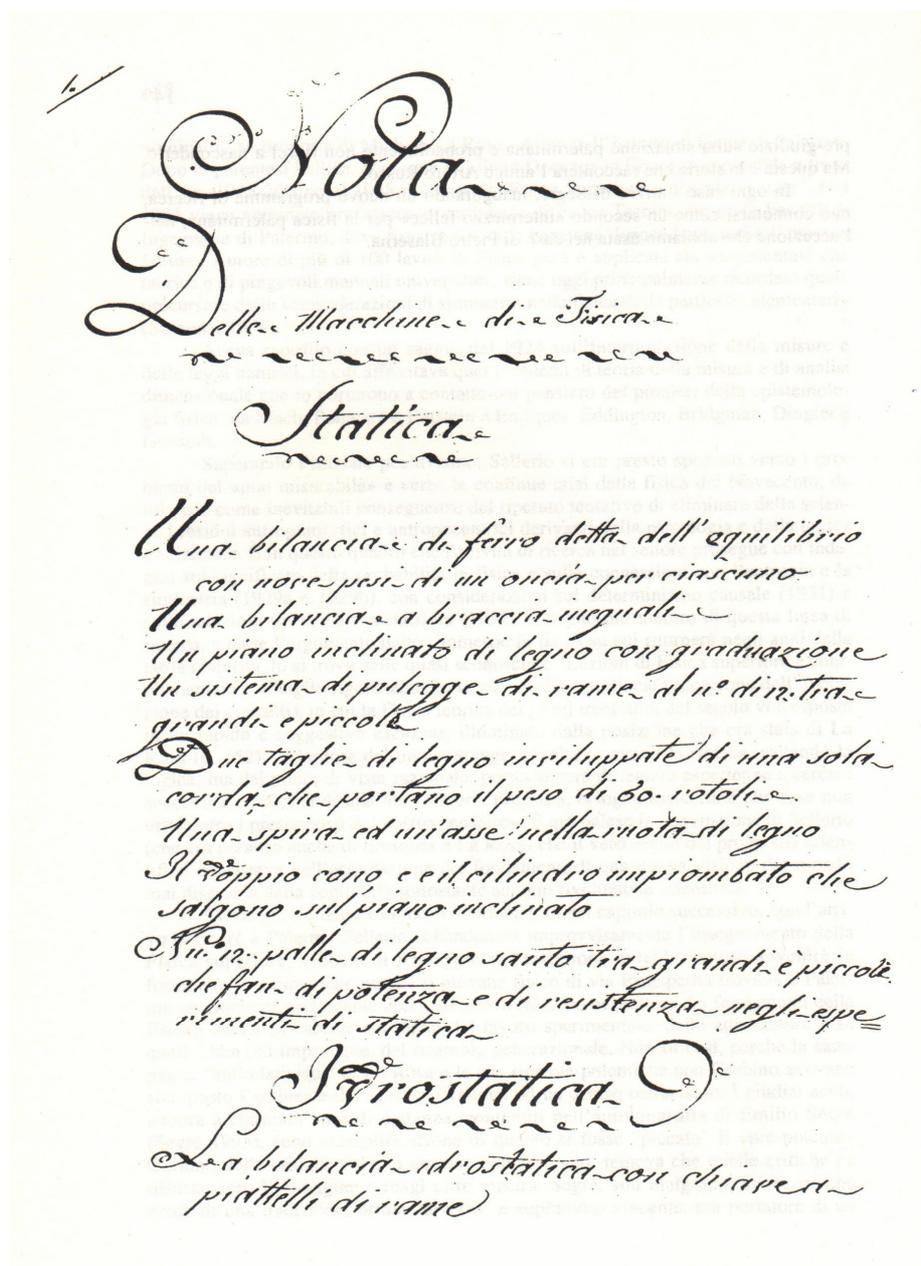

**Figura 1.** Copia della prima pagina dell'Inventario Scinà, tratto da [5].

A. Agliolo Gallitto *et al*, Due "paradossi meccanici" della Collezione Storica…

## 2. Doppio cono

Il più noto dei paradossi meccanici è certamente il doppio cono. Descritto per la prima volta da Leybourn nel 1694 [12], il doppio cono è formato dall'unione alla base di due coni identici. Quando il doppio cono è posizionato all'estremità inferiore di una coppia di binari inclinati e divergenti comincia spontaneamente a muoversi verso l'alto, dando l'impressione di sfidare la legge di gravitazione universale di Newton. A causa di questo comportamento, che contraddice il senso comune, questo strumento viene descritto come un paradosso meccanico. In figura 2 è mostrato il doppio cono della Collezione Storica degli Strumenti di Fisica dell'Università di Palermo (Inv. N.149).

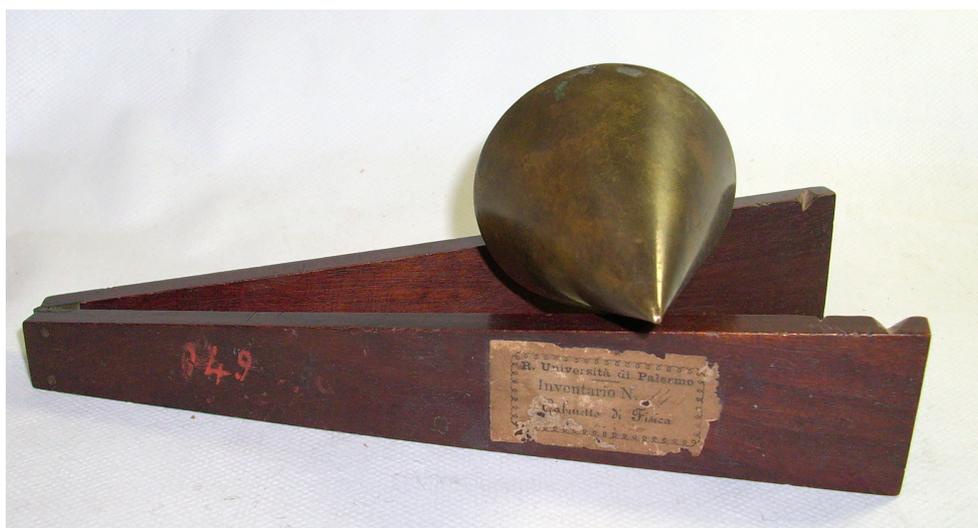

**Figura 2.** Il doppio cono della Collezione Storica degli Strumenti di Fisica (Inv. N. 149).

L'apparente paradosso è naturalmente spiegato nell'ambito delle leggi di Newton. Un corpo in presenza della forza di gravità si muove verso un minimo di energia potenziale gravitazionale. A causa della divergenza dei binari, durante il moto il centro di massa (CM) del doppio cono, che per simmetria si trova sull'asse di rotazione in corrispondenza del diametro massimo, si muove verso il basso mentre i punti di contatto con i binari si spostano verso l'apice di ogni cono. Come risultato complessivo, il CM si muove verso il basso in modo del tutto coerente con le leggi della meccanica di Newton [9,10,13].

Per capire se il doppio cono si muoverà verso l'alto, bisogna calcolare la variazione dell'altezza del suo asse rispetto al piano orizzontale dopo uno spostamento infinitesimo nella direzione orizzontale. Consideriamo un sistema di assi cartesiani con l'asse $x$ nella direzione orizzontale del moto, l'asse $y$ nel piano orizzontale (perpendicolare all'asse $x$) e l'asse $z$ nella direzione verticale. Figura 3 mostra una rappresentazione schematica della guida su cui si muove il doppio cono, dove abbiamo indicato con $\psi$ il semi angolo di apertura dei binari e con $\varphi$ il loro angolo di inclinazione, come indicato, rispettivamente, in figura 3(a) e 3(b). La rappresentazione schematica del doppio cono è mostrata in figura 4, dove abbiamo indicato con $\theta$ il semi angolo del cono, figura 4(a); con $\zeta$ l'asse perpendicolare alla guida e passante per il CM del doppio cono, figura 4(b).

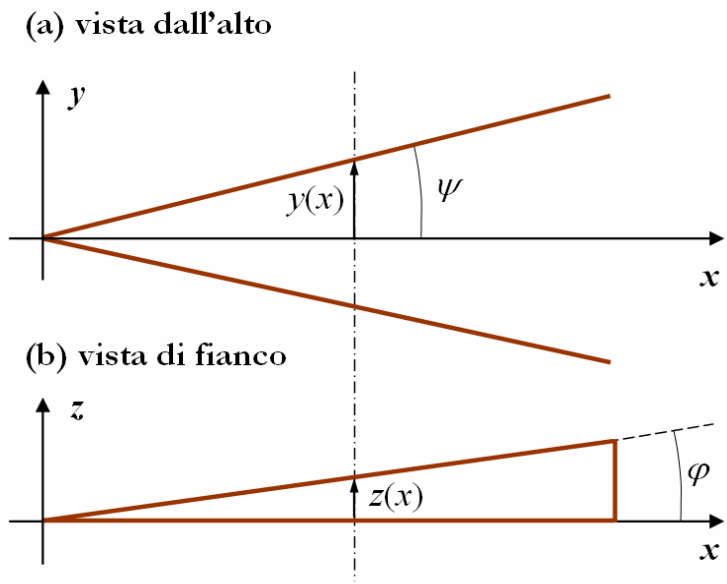

**Figura 3.** Schema dei binari su cui si muove il doppio cono.

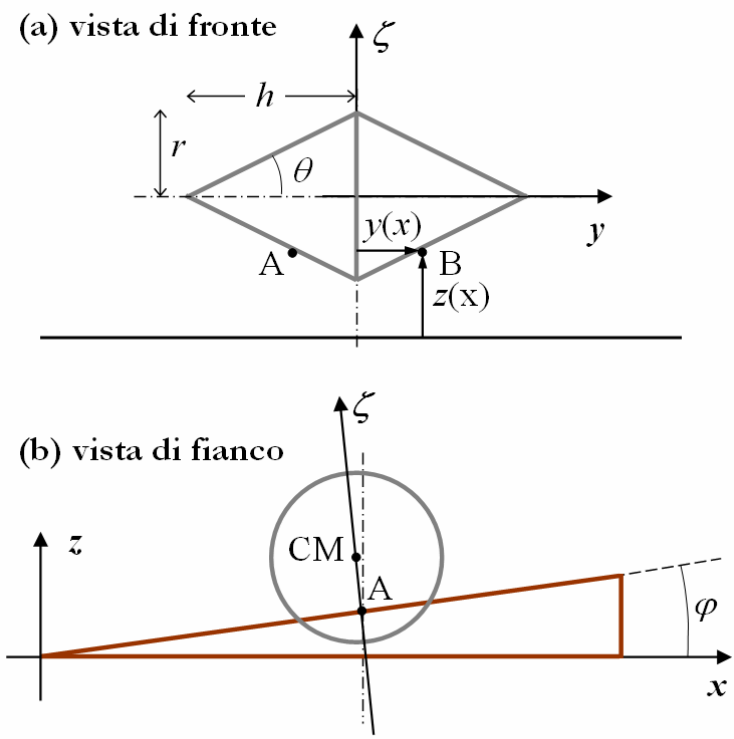

**Figura 4.** Schema del doppio cono sulla guida inclinata.

La distanza $y(x)$ tra i punti di contatto del doppio cono e i binari (indicati in figura 4(a) con A e B) aumenta all'aumentare dello spostamento $x$ nella direzione di moto dell'asse del doppio cono

$$dy = y(x + dx) - y(x) = \tan(\psi)\, dx \; ; \tag{1}$$

di conseguenza, la distanza $\zeta(y)$ tra l'asse del doppio cono e i punti di contatto con i binari diminuisce della seguente quantità

$$d\zeta = \zeta(y + dy) - \zeta(y) = \tan(\theta)\, dy = \tan(\theta)\tan(\psi)\, dx \; . \tag{2}$$

Poiché l'asse $\zeta$ è lungo la direzione perpendicolare ai binari, come mostrato in figura 4(b), esso forma un angolo $\varphi$ con la verticale. Pertanto, lo spostamento nella dirazione verticale $d\zeta'$ si ottiene moltiplicando $d\zeta$ per $\cos(\varphi)$

$$d\zeta' = \cos(\varphi)\, d\zeta = \cos(\varphi)\tan(\theta)\tan(\psi)\, dx \; . \tag{3}$$

I punti di contatto sui binari, i quali sono inclinati di un angolo $\varphi$ rispetto al piano orizzontale, all'aumentare di $x$ si innalzano della seguente quantità

$$dz = z(x + dx) - z(x) = \tan(\varphi)\, dx \; . \tag{4}$$

Lo spostamento netto in verticale dell'asse del doppio cono è dato da $dz - d\zeta'$. Per osservare un moto verso l'alto del doppio cono, deve quindi essere verificata la seguente condizione

$$dz - d\zeta' < 0 \;\;\Rightarrow\;\; \tan(\varphi) < \cos(\varphi)\tan(\theta)\tan(\psi) \; . \tag{5}$$

Usando la relazione trigonometrica tra il coseno e la tangente, $\cos(x) = \pm 1/[1 + \tan^2(x)]^{0.5}$, e risolvendo la disequazione (5) per $\tan(\varphi)$ otteniamo

$$\tan(\varphi) < \tan(\theta)\tan(\psi)/[1 - \tan^2(\theta)\tan^2(\psi)]^{0.5} \; . \tag{6}$$

Vale la pena notare che se la guida non è molto inclinata, e quindi l'angolo $\varphi$ è piccolo, $d\zeta' \approx d\zeta$ e la disequazione (6) si riduce a

$$\tan(\varphi) < \tan(\theta)\tan(\psi) \; . \tag{7}$$

Quando lo spostamento in verticale dell'asse del doppio cono è maggiore dello spostamento in verticale dei punti di contatto sulla guida, il CM del doppio cono scende e l'energia potenziale diminuisce, anche se i punti di contatto tra il doppio cono e la guida si muovono verso l'alto. Ciò dà luogo all'apparente paradosso che il doppio cono sale sul piano inclinato.

Per consentire ai visitatori di poter sperimentare personalmente il comportamento del doppio cono, abbiamo costruito una copia del doppio cono e della guida in legno, mostrati in figura 5.

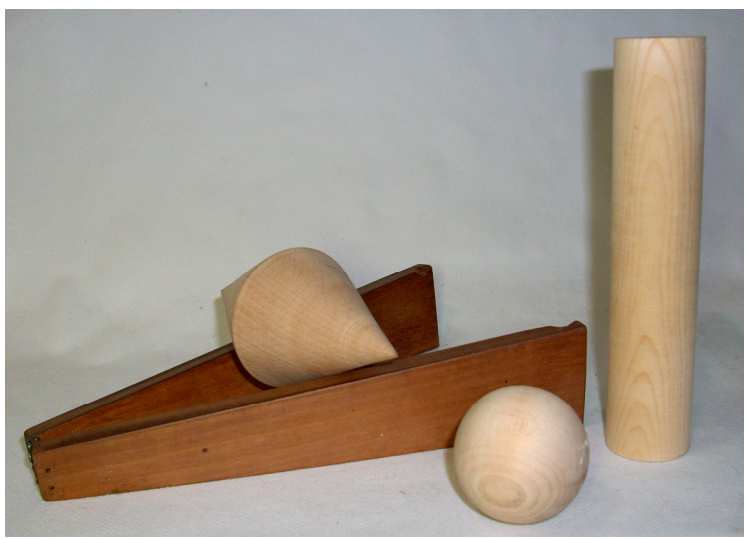

**Figura 5.** Guida di legno, doppio cono, cilindro e fera.

Le dimensioni del doppio cono sono $r = 30$ mm e $h = 65$ mm, con un angolo $\theta = \arctan(r/h) = 24.8°$. I binari sono inclinati di un angolo $\varphi = 6.5°$ e formano un semi angolo orizzontale $\psi = 15.3°$. Il doppio cono funziona abbastanza bene e ovviamente soddisfa la disequazione (6)

$$\tan(6.5°) < \tan(24.8°)\tan(15.3°))/[1 - \tan^2(24.8°)\tan^2(15.3°)]^{0.5} \quad \Rightarrow \quad 0.114 < 0.127 \qquad (8)$$

e la disequazione (7)

$$\tan(6.5°) < \tan(24.8°)\tan(15.3°) \quad \Rightarrow \quad 0.114 < 0.126 \,. \qquad (9)$$

Per una maggiore valenza didattica, il comportamento del doppio cono può essere confrontato con quello di un cilindro e quello di una sfera. Il cilindro presenta un comportamento regolare, cioè posizionato sopra la guida inclinata rotolerà verso il basso. La sfera invece si comporta come il cilindro quando è posta sopra la guida vicino al punto di convergenza dei binari e si comporta invece come il doppio cono quando è posta sopra la guida più lontano dal punto di convergenza. Quest'ultimo apparato, proposto nel 1996 da Martin Gardner [14], può essere usato per dimostrare sperimentalmente il teorema di Rolle [9].

Il teorema di Rolle afferma che data una funzione $U(x)$ con le seguenti proprietà:

- $U(x)$ è continua nell'intervallo chiuso $[x_1, x_2]$
- $U(x)$ è differenziabile ovunque nell'intervallo aperto $(x_1, x_2)$
- $U(x_1) = U(x_2)$

allora o $U(x)$ è costante oppure esiste un punto $x_0$ con $x_1 < x_0 < x_2$ in cui la derivata, rispetto a $x$, della funzione è nulla $dU(x)/dx = 0$.

Il teorema di Rolle ha un grande contenuto intuitivo e non verrà dimostrato in questa sede [15]; la sua applicazione al caso della sfera posta su due binari divergenti è particolarmente facile. La sfera poggia sul piano orizzontale sia nel punto convergenza dei due binari, che indichiamo con $x_1$, sia nel punto in cui i due binari si sono sufficientemente distanziati da permetterle di toccare il piano, che indichiamo con $x_2$; senza perdita di generalità, si può supporre che in questi due punti l'energia potenziale $U(x)$ della sfera sia nulla: $U(x_1) = U(x_2) = 0$. Orbene, l'energia potenziale $U(x)$ nell'intervallo $(x_1, x_2)$ è una funzione continua e derivabile (la sua derivata è proporzionale alla forza); essa non è costante e $U(x) > 0$ in $x_1 < x < x_2$, per cui esisterà un punto $x_0$ in cui la derivata di $U$ è nulla e $U(x_0)$ assume valore massimo. Ciò implica che $x_0$ è un punto di equilibrio instabile.

Il set di strumenti sopra descritti, completato con un pannello esplicativo in cui vengono illustrati i tre punti: cosa fare, cosa osservare e cosa accade, costituisce un interessante exhibit, cioè un elemento espositivo interattivo, che può essere usato durante le visite della Collezione per aumentare l'interesse degli studenti e del pubblico generico.

Nel prossimo paragrafo esamineremo il comportamento di un altro paradosso meccanico della Collezione, il cilindro impiombato, noto anche come il "cilindro disobbediente".

## 3. Cilindro impiombato

Quando il cilindro impiombato è posto sopra un piano inclinato, esso non rotola verso il basso come si aspetterebbe per un normale cilindro, ma rimane in equilibrio. A causa di questo comportamento apparentemente anomalo, il dispositivo è descritto come un paradosso meccanico [11]. Anche questo paradosso è presente nell'inventario Scinà [5]. Lo strumento della Collezione è mostrato in figura 6; esso non presenta alcuna indicazione del numero di inventario e pertanto potrebbe essere una copia di quello originale indicato nell'inventario Scinà. Tuttavia, considerato lo stato di ossidazione della zavorra di piombo, il disco potrebbe avere più di 100 anni.

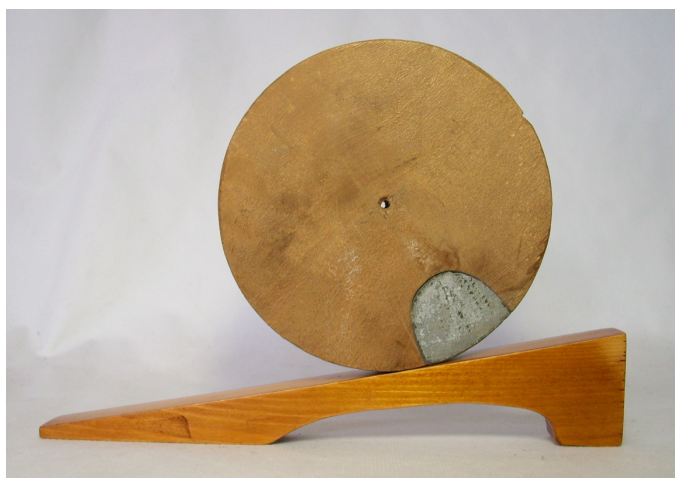

**Figura 6**. Il cilindro impiombato della Collezione, in equilibrio sopra il piano inclinato.

Il comportamento del cilindro impiombato è facilmente compreso considerando che esso nasconde una zavorra, generalmente di piombo, che sposta il centro di massa (CM) del sistema lontano dall'asse del cilindro. Il peso della zavorra crea un momento meccanico che consente al cilindro di risalire sul piano inclinato e fermarsi in una ben determinata posizione di equilibrio, in cui tutti i momenti meccanici esterni che agiscono sul cilindro si annullano.

Per calcolare l'angolo massimo del piano inclinato, per il quale il cilindro può rimanere in equilibrio statico, consideriamo un cilindro di raggio $R_c$ e densità $\rho_c$ avente un foro cilindrico di raggio $R_h$ con l'asse centrato a una distanza $d$ dall'asse del cilindro, come mostrato in figura 7.

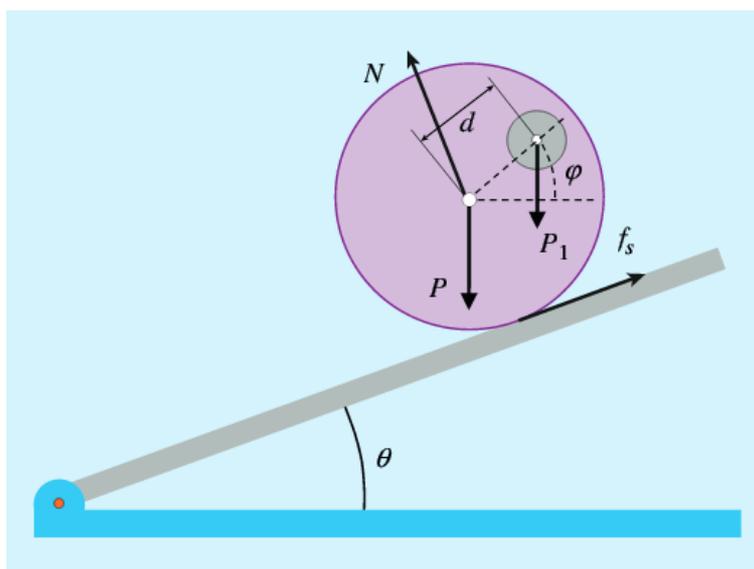

**Figura 7.** Rappresentazione del cilindro impiombato sul piano inclinato; immagine adattata da [11].

Nel foro è inserito un cilindretto di raggio $R_h$ e densità $\rho_h$, con $\rho_h \gg \rho_c$, per esempio un cilindretto di piombo, ottone o acciaio. Indichiamo con $P$ il peso del disco pieno (senza il foro), $P_1 = g\,(\rho_h - \rho_c)\,V_h$ il peso supplementare e $f_s$ la forza di attrito. Assumiamo anche che il coefficiente di attrito statico tra cilindro e piano sia sufficientemente grande da permettere al cilindro di rotolare senza scivolamento, quindi la forza di attrito statico equilibra sempre la componente della forza di gravità parallela al piano inclinato. Il cilindro rotola in salita sul piano inclinato, se il momento di $P_1$ rispetto al punto di contatto è maggiore del momento di $P$ rispetto allo stesso punto (vedi figura 7). Si raggiunge una posizione di equilibrio quando questi due momenti sono uguali e opposti.

Considerando una generica posizione del cilindro sul piano inclinato, è facile dimostrare che il cilindro rimane in equilibrio quando la somma dei momenti rispetto al punto di contatto del disco col piano inclinato è nulla

$$P\,R_c \sin(\theta) - P_1\,[d \cos(\varphi) - R_c \sin(\theta)] = 0 \,, \tag{10}$$

da cui si ottiene

$$R_c \sin(\theta) = R_{cm} \cos(\varphi) \,, \tag{11}$$

dove $R_{cm} = P_1\,d\,/\,(P + P_1)$ è il raggio del CM del sistema. Per un angolo fisso $\theta$ del piano inclinato, ci sono due posizioni di equilibrio in corrispondenza degli angoli $+\varphi$ e $-\varphi$, che corrispondono rispettivamente alla posizione di massimo e minimo di energia potenziale gravitazionale. Aumentando l'angolo $\theta$, si ottiene l'angolo massimo, $\theta_{max}$, quando $\varphi = 0$, cioè quando l'asse del cilindro e l'asse del foro giacciono su una linea orizzontale.

All'aumentare dell'angolo $\theta$ del piano inclinato, se il coefficiente di attrito statico è sufficientemente grande, il cilindro può rimanere in equilibrio fino all'angolo massimo $\theta_{max}$, superato il quale il cilindro inizia a rotolare verso il basso. La posizione limite corrisponde alla condizione in cui l'asse del cilindro e quello della zavorra giacciono sulla stessa linea orizzontale, mentre il CM del cilindro e il punto di contatto giacciono sulla stessa linea verticale. Questo risultato permette di determinare sperimentalmente la posizione del CM del sistema.

Poiché il CM del cilindro dista $R_{cm}$ dall'asse del cilindro, quando esso è posto sul piano inclinato di un angolo $\theta < \theta_{max}$, in una opportuna posizione, esso può rotolare verso l'alto sul piano inclinato. In figura 8 è riportata un'illustrazione del moto del cilindro impiombato sopra il piano inclinato tratta da [16]. Il cilindro si fermerà nella posizione di equilibrio, in cui l'energia potenziale è minima. Rotolando verso l'alto sopra il piano inclinato, l'altezza del CM diminuirà, in accordo con le leggi della Fisica.

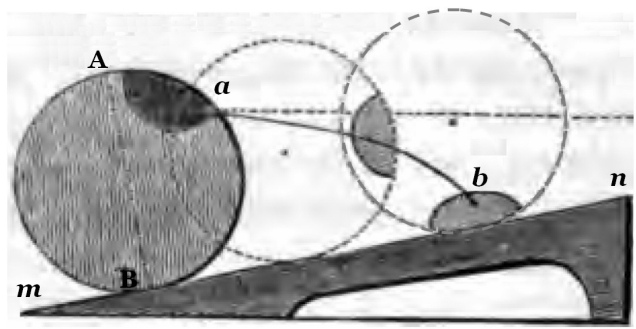

**Figura 8.** Illustrazione del moto del cilindro impiombato sul piano inclinato, immagine adattata da [16].

Il cilindro impiombato della Collezione, a causa della forma irregolare della zavorra di piombo, non si presta a essere usato per scopi didattici. Per questo motivo, abbiamo realizzato un cilindro di legno di raggio $R_c = 45$ mm e densità $\rho_c = 0.5$ kg/m³ con due fori di raggio $R_h = 16$ mm, uno coassiale con il cilindro e l'altro decentrato di una distanza $0.5R_c$. Abbiamo quindi realizzato due cilindretti di raggio $R_h$, uno di legno e l'altro di ottone ($\rho_h = 8.5$ kg/m³), da usare come zavorra. Inserendo il cilindretto di ottone nel foro centrale e quello di legno nel foro laterale il sistema si comporta in modo normale. Viceversa, inserendo il cilindretto di legno nel foro centrale e quello di ottone nel foro decentrato il sistema rimane in equilibrio sopra il piano inclinato, manifestando un comportamento apparentemente paradossale. L'angolo massimo del piano inclinato per il quale il cilindro rimane in equilibrio può essere calcolato facilmente dall'equazione (11) per $\varphi = 0$

$$\theta_{max} = \arcsin(R_{cm} / R_c) \approx 10° \; . \tag{12}$$

Il kit così realizzato costituisce un interessante exhibit per lo studio delle proprietà del centro di massa.

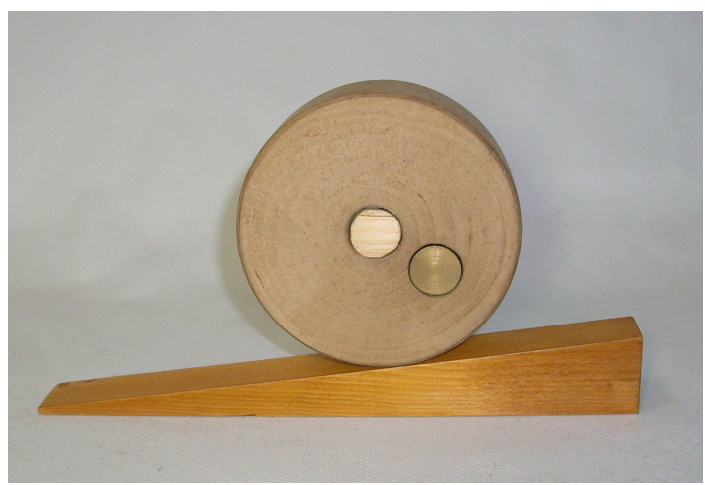

**Figura 9.** Cilindro impiombato con zavorra di ottone.

## 4. Discussione e conclusioni

Le collezioni storiche di strumenti scientifici sono importanti per molti aspetti. Dal punto di vista storico, gli strumenti delle collezioni consentono di ripetere esperimenti che hanno contribuito significativamente allo sviluppo delle scienze, mentre dal punto di vista didattico sono una fonte di idee per realizzare apparati sperimentali ed exhibit espositivi di elevato contenuto scientifico. Per esempio, gli esperimenti con il doppio cono e il cilindro impiombato possono essere usati proficuamente in attività laboratoriali, in quanto gli studenti all'inizio dello studio della meccanica sono attratti dai paradossi e aiutati dal docente spiegano i fenomeni con le loro conoscenze sulle leggi di Newton. Questi paradossi meccanici, insieme ai giochi scientifici, possono dunque essere usati per aumentare l'attenzione degli studenti durante le lezioni e, in generale, aumentare l'interesse per lo studio della Fisica.

In conclusione, abbiamo analizzato e discusso due paradossi meccanici, il doppio cono e il cilindro impiombato, della Collezione Storica degli Strumenti di Fisica dell'Università di Palermo. Abbiamo determinato le condizioni geometriche a cui deve soddisfare il doppio cono per risalire sulla guida inclinata, determinato le condizioni di equilibrio statico del cilindro impiombato sopra il piano inclinato e la massima inclinazione che può avere il piano affinché il cilindro rimanga in equilibrio. Abbiamo quindi visto che la Collezione Storica degli Strumenti di Fisica può essere una risorsa da cui prendere idee per sviluppare attività laboratoriali ed exhibit rivolti a studenti e pubblico generico, al fine di aumentare l'interesse degli studenti verso lo studio della Fisica.